\documentclass[a4paper]{cas-sc}

\usepackage[numbers]{natbib}

\usepackage{amsthm}
\usepackage[T1]{fontenc}
\usepackage{caption}
\usepackage{graphicx}
\usepackage{float} 
\usepackage{subcaption}
\usepackage{multirow}
\usepackage{algorithm}
\usepackage{algpseudocode}
\usepackage{bm}
\usepackage{amsfonts}
\usepackage{amsmath}
\usepackage{array}

\def\tsc#1{\csdef{#1}{\textsc{\lowercase{#1}}\xspace}}
\tsc{WGM}
\tsc{QE}
\tsc{EP}
\tsc{PMS}
\tsc{BEC}
\tsc{DE}
\begin{document}
\let\WriteBookmarks\relax
\def\floatpagepagefraction{1}
\def\textpagefraction{.001}

\shorttitle{The Impact of Misclassifying Non-confounding Covariates as Confounders on the Causal Inference}

\shortauthors{Yonghe Zhao et~al.}

\title [mode = title]{Does Misclassifying Non-confounding Covariates as Confounders Affect the Causal Inference within the Potential Outcomes Framework?}  

%A Study on the Impact of Non-confounding Covariates on the Inferential Performance of Methods based on the Potential Outcome Framework

%A study on the impact of misclassifying non-confounding covariates as confounders on causal inference within the potential outcomes framework.

% Does Misclassifying Non-confounding Covariates as Confounders Affect the Causal Inference within the Potential Outcomes Framework?

%Does the Presence of Non-confounding Covariates Affect the Inferential Performance of Methods based on the Potential Outcome Framework?
% \tnotemark[1,2]

% \tnotetext[1]{This document is the results of the research
%    project funded by the National Science Foundation.}

% \tnotetext[2]{The second title footnote which is a longer text matter
%    to fill through the whole text width and overflow into
%    another line in the footnotes area of the first page.}

\author[1]{Yonghe Zhao}[orcid=0000-0003-2613-7526]

%\cormark[1]


\ead{yhzhao21@mails.jlu.edu.cn}

%\ead[url]{www.cvr.cc, cvr@sayahna.org}

% \credit{Conceptualization, Formal analysis, Methodology, Writing - original draft}

\affiliation[1]{organization={School of Artificial Intelligence, Jilin University},
    addressline={Qianjin Street 2699}, 
    city={Changchun},
    % citysep={}, % Uncomment if no comma needed between city and postcode
    postcode={130012}, 
    state={Jilin},
    country={China}}

\author[1]{Qiang Huang}[style=chinese]

% \ead{huangqiang18@mails.jlu.edu.cn}

% \credit{Methodology, Writing - review \& editing}

% \author[1]{Siwei Wu}[style=chinese]

% \ead{wusw22@mails.jlu.edu.cn}

% \credit{Data curation, Software}

\author[2]{Shuai Fu}[style=chinese]

% \ead{fushuai@buaa.edu.cn}

% \credit{Writing - review \& editing}

% \affiliation[2]{organization={Department of Data Analysis, Baidu},
%     addressline={Shangdi 10th Street 10}, 
%     city={Beijing},
    % citysep={}, % Uncomment if no comma needed between city and postcode
%     postcode={100085}, 
    % state={Jilin},
%     country={China}}

\affiliation[2]{organization={State Key Laboratory of Software Development Environment, Beihang University},
    addressline={Xueyuan Road 37}, 
    city={Beijing},
    % citysep={}, % Uncomment if no comma needed between city and postcode
    postcode={100191}, 
    % state={Beijing},
    country={China}}

\author[1]{Huiyan Sun}[style=chinese]
% [orcid=0000-0002-4664-7147]
\cormark[1]
% \fnmark[1,3]
% \ead{huiyansun@jlu.edu.cn}
% \ead[URL]{https://www.researchgate.net/profile/Huiyan-Sun}

% \credit{Writing - review \& editing, Supervision, Validation}

%\affiliation[2]{organization={Sayahna Foundation},
%    city={Jagathy},
    % citysep={}, % Uncomment if no comma needed between city and postcode
%    postcode={695014}, 
%    state={Trivandrum},
%    country={India}}

%\author%
%[1,3]
%{Rishi T.}
%\cormark[2]
%\fnmark[1,3]
%\ead{rishi@stmdocs.in}
%\ead[URL]{www.stmdocs.in}

%\affiliation[3]{organization={STM Document Engineering Pvt Ltd.},
%    addressline={Mepukada}, 
%    city={Malayinkil},
    % citysep={}, % Uncomment if no comma needed between city and postcode
%    postcode={695571}, 
%    state={Trivandrum},
%    country={India}}

\cortext[cor1]{Corresponding author}
%\cortext[cor2]{Principal corresponding author}

%\fntext[fn1]{This is the first author footnote. but is common to third
%  author as well.}
%\fntext[fn2]{Another author footnote, this is a very long footnote and
%  it should be a really long footnote. But this footnote is not yet
%  sufficiently long enough to make two lines of footnote text.}

%\nonumnote{This note has no numbers. In this work we demonstrate $a_b$
%  the formation Y\_1 of a new type of polariton on the interface
%  between a cuprous oxide slab and a polystyrene micro-sphere placed
%  on the slab.
%  }

\begin{abstract}
The Potential Outcome Framework (POF) plays a prominent role in the field of causal inference. Most causal inference models based on the POF (CIMs-POF) are designed for eliminating confounding bias and default to an underlying assumption of Confounding Covariates. This assumption posits that the covariates consist solely of confounders. However, the assumption of Confounding Covariates is challenging to maintain in practice, particularly when dealing with high-dimensional covariates. While certain methods have been proposed to differentiate the distinct components of covariates prior to conducting causal inference, the consequences of treating non-confounding covariates as confounders remain unclear. This ambiguity poses a potential risk when conducting causal inference in practical scenarios. In this paper, we present a unified graphical framework for the CIMs-POF, which greatly enhances the comprehension of these models' underlying principles. Using this graphical framework, we quantitatively analyze the extent to which the inference performance of CIMs-POF is influenced when incorporating various types of non-confounding covariates, such as instrumental variables, mediators, colliders, and adjustment variables. The key findings are: in the task of eliminating confounding bias, the optimal scenario is for the covariates to exclusively encompass confounders; in the subsequent task of inferring counterfactual outcomes, the adjustment variables contribute to more accurate inferences. Furthermore, extensive experiments conducted on synthetic datasets consistently validate these theoretical conclusions.
\end{abstract}

% \includegraphics{figs/grabs.pdf}

% \begin{highlights}
% \item Causality
% \item Counterfactual Inference
% \item Explanable Artificial Intelligence
% \end{highlights}

% Keywords
\begin{keywords}
Causal Inference \sep Non-Confounding Covariates \sep Post-treatment Variables \sep Confounders
\end{keywords}

\maketitle

\section{Introduction}
In recent years, causal inference has attracted increasing attention across various domains, including epidemiology, healthcare, and economics\cite{NIPS2017_6a508a60,2016Causal,johansson2018learning}, etc. Compared to correlation, causality represents a more fundamental relationship between variables, revealing the directionality and determinacy\cite{imbens2015causal}. Randomized controlled trials (RCTs) are widely regarded as an effective means of exploring causality\cite{JudeaCausality}. However RCTs are time-consuming and expensive, even involving ethical issues in some scenarios\cite{2010Reputation,2016Assessing,2011Unexpected}. How to conduct causal inference directly from collected observational data is a research topic of widespread concern.

Different from the randomness of treatments in RCTs, the main challenge of causal inference in observational studies is the unknown mechanism of treatment assignment\cite{imbens2015causal}. That is, there may be various deviations in observational data,  such as confounders, covariates that influence both treatment and outcome variables. Various causal inference methods for de-confounding of observed confounders are proposed based on the POF, denoted as CIMs-POF, such as reweighting\cite{rosenbaum1983central,2019Robust,lee2011weight,Austin2011An,imai2014covariate}, matching\cite{L2021Combining,2017Informative,JMLR21}, causal trees\cite{2010BART,Hill2011Bayesian,2017Estimation}, confounding balanced representation learning\cite{2018LearningWeighted,johansson2018learning,schwab2019perfect,2019AdversarialDu}, etc. These methods typically rely on three foundational assumptions in the POF\cite{imbens2015causal}: Stable Unit Ttreatment Value (SUTV), Unconfoundedness, and Positivity assumption. Moreover, most CIMs-POF are commonly assumed that there exists a fourth underlying hypothesis named as Confounding Covariates\cite{kuang2017treatment}. This hypothesis posits that all covariates act as confounders.
Several CIMs-POF have relaxed the assumption of Confounding Covariates and incorporated pre-treatment variables in covariates. The pre-treatment variables include instrumental, confounding, and adjustment variables that are not affected by the treatment variables\cite{Rubin08}. However, ensuring the fulfillment of this assumption remains elusive in practice\cite{ding2015adjust,vanderweele2011new}, particularly in scenarios involving high-dimensional covariates.

The implementation of causal inference faces significant challenges in real-world scenarios due to the inability to adequately guarantee the assumption of Confounding Covariates, making it a topic of considerable attention and concern.
Jessica et al.\cite{myers2011effects} validated that conditioning on instrumental variables introduces significant estimation error through numerical experiments. Additionally, they emphasize the importance of distinguishing between confounding and instrumental variables.
The ${\rm D^{2}VD}$ model\cite{kuang2017treatment} was proposed as a data-driven approach to identify confounding and adjustment variables. In line with this, Tyler et al.\cite{vanderweele2019principles} advocated first identifying the causes of the treatment and outcome variables among the covariates before conducting causal inference\cite{vanderweele2019principles}. Essentially, these causes are equivalent to the pre-treatment variables.
Furthermore, Negar et al. devised a DR-CFR model\cite{hassanpour2019learning} that utilizes representation learning to obtain distinct representations of confounding, instrumental, and adjustment variables. Building upon this work, Kun Kuang et al. proposed the DeR-CFR model\cite{wu2020learning}, which extends the DR-CFR model by imposing additional constraints on instrumental variables to enhance the differentiation of these pre-treatment variables.
However, these methods have only studied one or a few components of the covariates. Specifically, there is limited literature on the estimation error of CIMs-POF caused by conditioning the post-treatment variables, such as mediators and collider variables. When post-treatment variables are present, their interactions with pre-treatment variables may compromise the effectiveness of the aforementioned methods\cite{2000Causality}. There is an urgent need for a systematic examination and empirical evidence regarding the consequences of mistakenly identifying various non-confounding variables as confounding ones for causal inference.

In response to the aforementioned research gap, we conduct a comprehensive examination of the impact of different non-confounding covariates, encompassing pre-treatment and post-treatment variables, on the performance of causal inference within POF.
Drawing on a unified graphical framework proposed in this paper for the CIMs-POF, we quantify the theoretical impact and substantiates it with empirical evidence. Furthermore, we provide practical guidelines for the optimal implementation of the CIMs-POF. The main contributions of this paper are outlined as follows.
\begin{itemize}
\item The paper presents clear conclusions from both theoretical and experimental perspectives regarding the impact of mistakenly treating non-confounding covariates as confounding ones on the causal inference. These findings provide valuable guidance for the practical application of CIMs-POF.
\item We develop a unified graphical framework for the CIMs-POF, which provides significant advantages in comprehensively understanding their underlying principles. Based on this graphical framework, we quantitatively evaluate, from a theoretical perspective, the estimation errors of the CIMs-POF in estimating treatment effects in the presence of various non-confounding covariates.
\item We conduct extensive experiments on synthetic datasets, and the experimental results are consistent with the general conclusions derived from theoretical analysis. Furthermore, we analyze the practical nuances of specific non-confounding covariates, such as colliders and the outcome-influenced variables.
\end{itemize}
\section{Related Works}
\subsection{Causal Inference Methods based on POF}
\begin{figure}[ht]
\centering
\includegraphics[width=0.95\textwidth]{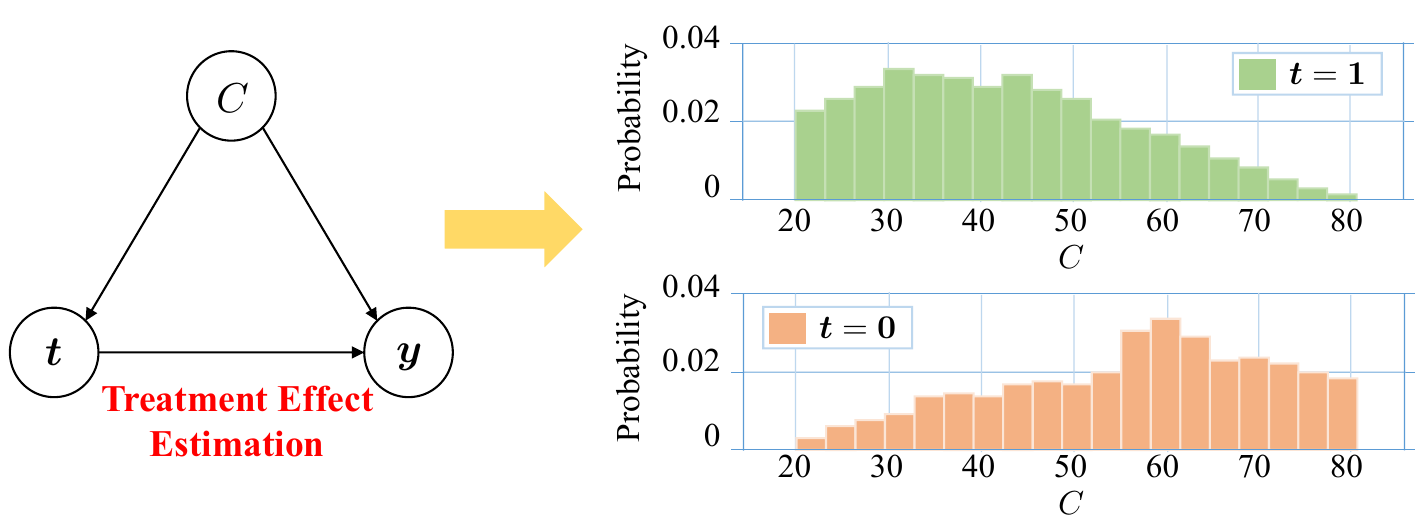}
\caption{The issues engendered by confounding covariates: inconsistent distribution of $C$ amidst discrete $t$ values.}
\label{confounders}
\vspace{-0.3cm}
\end{figure}

The POF\cite{rubin1974estimating,splawa1990application}, also known as the Rubin Causal Model, is not a specific method but rather a framework for conducting causal inference on observational data. Based on the aforementioned three assumptions, its main objective is to address the challenge of confounding bias in causal inference. As shown in Fig. \ref{confounders}, the confounding covariates $C$ affect the selection of treatment $t$ thus leading to inconsistent distribution of $C$ among discrete $t$ values. Consequently, this phenomenon results in inaccurate counterfactual inference, which is similar to the domain adaptation problem\cite{yao2020survey}.

Therefore, the main objective of CIMs-POF is to achieve a balanced distribution of confounding covariates among different treatment groups\cite{yao2020survey}. These methods can be categorized into three categories based on the spatial domains: subspace of samples, sample space, and feature space. 
Methods based on the subspace of samples divide the samples based on certain balanced measures and then evaluate the causal effects in approximately balanced subspaces. Representative methods include Stratification\cite{2015Principal,hullsiek2002propensity}, Matching\cite{L2021Combining,2017Informative,JMLR21}, and the subspace partitioning methods based on decision trees, such as Bayesian Additive Regression Trees (BART)\cite{2010BART} and Random Causal Forest (RCF)\cite{2017Estimation}. These methods rely on the assumption of Positivity. 
Different from acquiring balanced subspaces, methods based on the sample space make weighted adjustments to the observational data of the original sample space to eliminate the influence of the confounders. These methods mainly include Inverse Propensity Weighting (IPW)\cite{rosenbaum1983central}, Doubly Robust estimator (DR)\cite{2019Robust}, Boosting Algorithm for Estimating Generalized Propensity Scores\cite{zhu2015boosting}, Covariate Balancing Propensity Score (CBPS)\cite{imai2014covariate}, and so on. Weight-based adjustment methods depend on the rationality of constructing weights. 
Methods based on the feature space aim to acquire a balanced representation of the confounding covariates in an abstract representation space, which is crucial for performing downstream counterfactual inference tasks. Notable methods include the Treatment-Agnostic Representation Network (TARNet)\cite{shalit2017estimating}, Counterfactual Regression (CFR)\cite{shalit2017estimating}, and Local Similarity Preserved Individual Treatment Effect (SITE)\cite{yao2018representation}.
\subsection{Methods for Identifying Complex Covariate Components}
The aforementioned CIMs-POF assume the assumption of Confounding Covariates. However, practical considerations surrounding covariates often involve complex components that encompass both confounding and non-confounding variables. This has sparked significant research interest in addressing these issues. 
Jessica et al.\cite{myers2011effects} presented the results of two simulation studies aimed at demonstrating that treatment effect estimate, conditioned on a perfect instrumental variable (IV) or a near-IV, may exhibit greater bias and variance compared to the unconditional estimate\cite{myers2011effects}. 
Kun Kuang et al. proposed a Data-Driven Variable Decomposition (${\rm D^{2}VD}$) algorithm\cite{kuang2017treatment} which automatically separate confounders and adjustment variables to estimate treatment effect more accurately. 
Tyler et al. put forward a practical approach\cite{vanderweele2019principles} for making confounder selection decisions based on the availability of knowledge regarding whether each covariate is a cause of the treatment or outcome variables. 
Negar et al. proposed the Disentangled Representations for Counterfactual Regression (DR-CFR) algorithm\cite{hassanpour2019learning} to identify disentangled representations of instrumental, confounding, and adjustment variables. Specifically, the DR-CFR algorithm constructs an objective function that relies on the independence between the representation of instrumental variables and the representation of confounding and adjustment variables. 
On this basis, Kun Kuang et al. constructed the Decomposed Representations for Counterfactual Regression (DeR-CFR) model\cite{wu2020learning}. This model introduces conditional independence constraints between the representation of instrumental variables and outcome variables to enhance the objective function of the DR-CFR algorithm. 
However, these methods have not comprehensively investigated the estimation error caused by non-confounding covariates in causal inference, such as post-treatment variables.
\section{The Impact of Misclassifying Non-confounding Covariates as Confounders on Causal Inference}
\subsection{Problem Setting}
\begin{figure}[ht]
\centering
\includegraphics[width=0.65\textwidth]{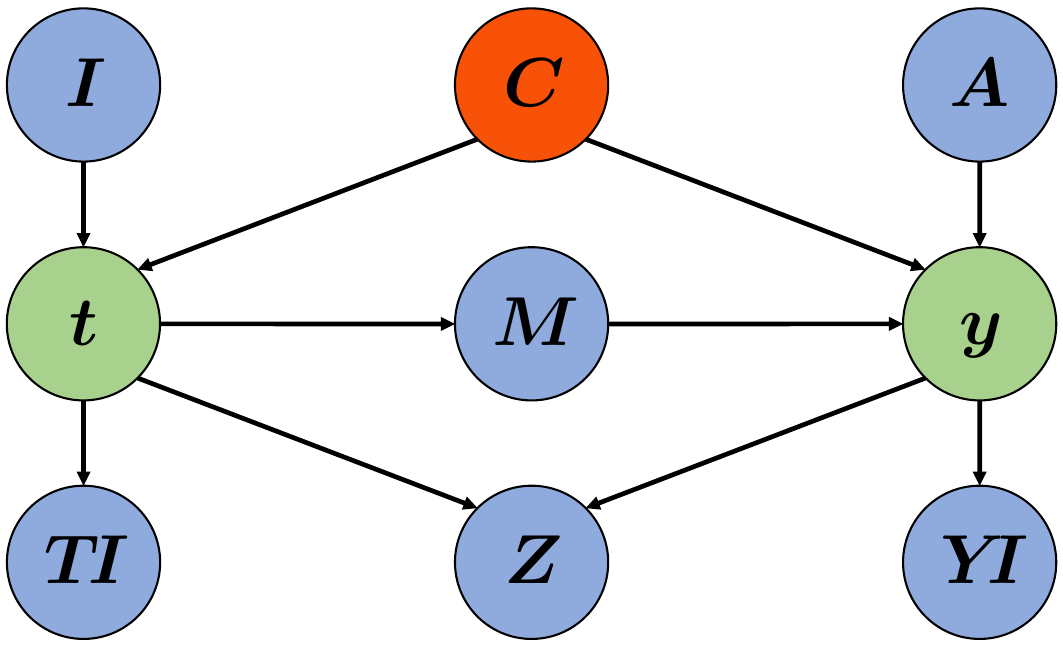}
\caption{The Causal Graph within the problem setting. The variables in the blue background represent potential covariates $X$.}
\label{CASAUL}
\vspace{-0.3cm}
\end{figure}

The CIMs-POF investigated in this paper require the fulfillment of four assumptions\cite{imbens2015causal,kuang2017treatment}: SUTV, Unconfoundedness, Positivity, and Confounding Covariates.
Additionally, the CIMs-POF primarily aim at observational data, represented as $\mathcal{D} = \{X_{i},t_{i},y_{i}\}_{i = 1}^{N}$, which encompasses $N$ independent and identically distributed samples. Specifically, as depicted in Fig. \ref{CASAUL}, we assume that covariates $X$ may include confounder $C$, mediator $M$, collider $Z$, instrumental variable $I$, adjustment variable $A$, treatment-influenced variable $\textit{TI}$, and the outcome-influenced variable $\textit{YI}$. The specific nature of these covariates needs further determination in practical applications.
The treatment variable $t \in \{0,1\}$ represents the control group ($t=0$) or the treatment group ($t=1$). Each value of the treatment variable, denoted as $t_{i}$, corresponds to a potential outcome indicated by $y(t_{i})$. These potential outcomes can be further categorized into the factual outcome $y^{f}$ and the counterfactual outcome $y^{cf}$. Within the dataset $\mathcal{D}$, for the $i$-th sample, only the factual outcome $y^{f}(t_{i})$ corresponding to $t_{i}$ is accessible, while the other counterfactual outcomes $y^{cf}(1-t_{i})$ remain unobserved. The primary objective of these CIMs-POF is to infer $y^{cf}(1-t_{i})$ and assess the treatment effects. This paper aims to quantify the specific impact of various non-confounding covariates on the performance of these methods in evaluating treatment effects, which is referred to as the Impact of Non-Confounding Covariates (INCC).
%\subsection{Assumptions} 
%Referring to the  basic of the POF, the CIMs-POF necessitate the fulfillment of three assumptions: stable unit treatment value (SUTV), unconfoundedness, and positivity assumption\cite{imbens2015causal}.
%
%The SUTV assumption includes: firstly, the potential outcome of each individual is not affected by the treatment of any other individual, in other words, individuals are independent; secondly, there is no measurement error in the factual observational outcome. 
%
%The unconfoundedness assumption represents that the treatment variable is independent of the outcome variable given the covariates $X$, i.e., $\mathcal{T}\perp\mathcal{Y}|X$. With this unconfoundedness assumption, for the samples with the same covariates $X$, their treatment assignment can be viewed as random.
%
%The positivity assumption, commonly referred to as the overlap assumption, posits that each value of $X$ can be assigned to any treatment with a non-zero probability, specifically $p(t|X=x) >0, \forall\ t \in \mathcal{T}, x \in X$. The purpose of counterfactual inference is to assess differences across treatments, and the model is meaningless if some treatments can not be observed or are not meaningful. 
%
%In addition, the CIMs-POF concerned in this paper make an underlying assumption: Confounding Covariates, i.e., all covariates are confounders.
\subsection{A Unified Graphical Framework of the CIMs-POF}
\begin{figure}[!ht]
\centering
\includegraphics[width=0.8\textwidth]{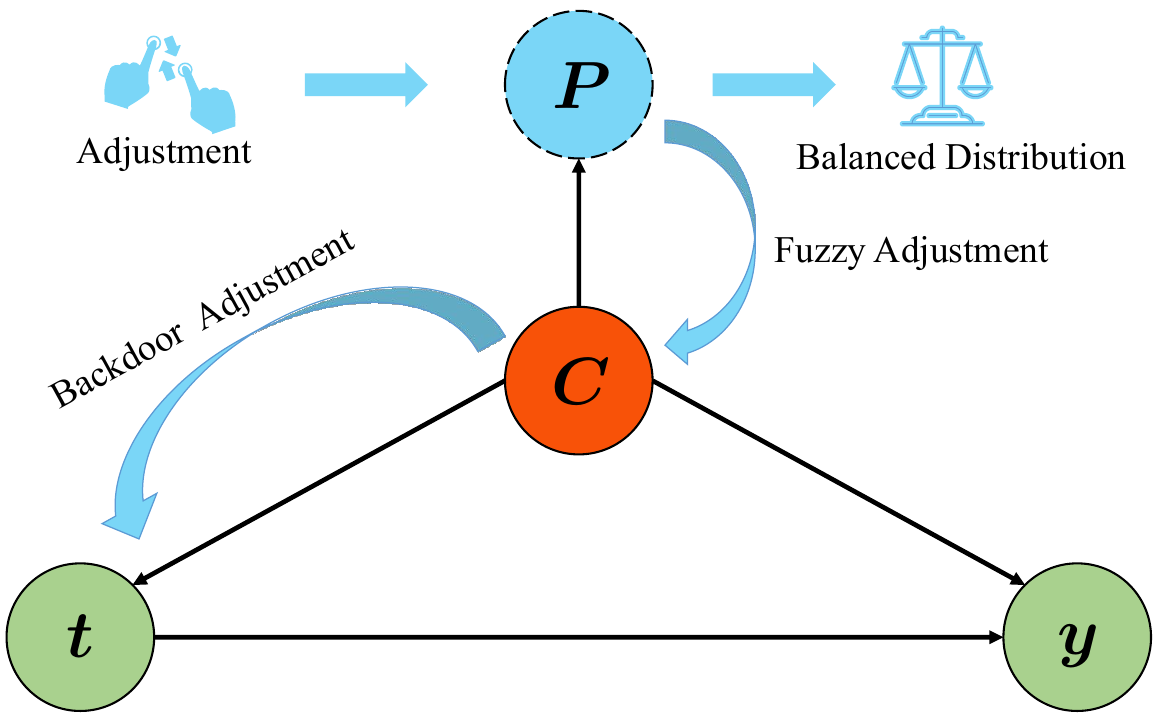}
\caption{A Unified Graphical Framework of the CIMs-POF: Adjusting for confounding variable $C$ through controlling $P$ is a fuzzy adjustment, corresponding to the backdoor adjustment in Structural Causal Model.
}
\label{UnifiedPers}
%\vspace{-0.3cm}
\end{figure}

To comprehensively analyze the INCC, we propose a unified graphical framework, as illustrated in Fig. \ref{UnifiedPers}. According to the assumption of the Unconfoundedness and Confounding Covariates, the main purpose of these CIMs-POF is to eliminate the observed confounding bias. In the framework depicted in Fig. \ref{UnifiedPers}, we introduce a method called Fuzzy Adjustment by Proxy (FAP) to eliminate confounding bias. This method involves two steps: firstly, establishing a proxy mapping denoted as $P = f(C)$, which captures the confounding covariates such as propensity scores; and secondly, indirectly adjusting the confounding bias by modifying the distribution of the proxy variable $P$ across discrete values of $t$.

To understand this graphical framework, we present specific instantiations of the proxy variable $P$ in various CIMs-POF, along with their corresponding adjustment approaches, in Table~\ref{table:1}. The Table~\ref{table:1} showcases representative CIMs-POF, which encompass different methods operating within the subspace of samples, such as Stratification, Matching, and RCF. Additionally, it includes methods like IPW and CBPS that are based on the original sample space, as well as CFR, a model in the representation space.
We illustrate the propensity score matching method as an example. This method involves initially estimating the propensity score, denoted as $P = P(t = 1|C)$, which represents the probability of receiving treatment given the confounding variables. Subsequently, samples with similar propensity scores are matched between the treatment and control groups to estimate the treatment effects. While the remaining CIMs-POF employ different instantiations of the proxy $P$ and corresponding adjustment approaches, the underlying graphical framework remains consistent. This consistency is depicted in Figure \ref{UnifiedPers}, which visually highlights the shared elements among the different CIMs-POF. 
\begin{table}[!t]
\centering
\caption{The instantiations of various CIMs-POF regarding the unified graphical framework.}
\label{table:1}
%\resizebox{1\columnwidth}{!}{
\setlength{\tabcolsep}{0.05\textwidth}{ %set colWidth
\begin{tabular}{l|c|c}
\hline
 Method & Instantiation of Proxy $P=f(C)$ &  Adjustment Approach \\
 \hline
 Stratification & Propensity Scores & Stratified-weighted average  \\ 
 Matching & Propensity Scores & $1-1/1-N$ matching  \\ 
 RCF & Leaf Scores & Leaf-weighted average  \\ 
 IPW & Propensity Scores & Reweighting  \\ 
 CBPS & Distribution properties/Propensity score & Reweighting  \\ 
 CFR & Representation Network & Network Parameters Learning  \\ 
\hline
\end{tabular}
}
%}
\end{table}

The backdoor adjustment criterion\cite{JudeaCausality,pearl2012class} in Structural Causal Models is fundamentally aligned with the FAP. The key distinction lies in the direct and precise adjustment of confounders $C$ through backdoor adjustment, as opposed to the FAP, which employs an indirect and fuzzy adjustment of the proxy variable $P$. In scenarios where the confounding covariates comprise high-dimensional continuous variables, performing exact adjustment using the backdoor criterion becomes computationally challenging. However, the FAP offers an effective solution by reducing computational complexity and achieving an approximate adjustment effect through fuzzy control of the low-dimensional proxy variable $P$.
\subsection{The Performance of FAP in Eliminating Confounding Bias}
To investigate the mechanisms by which non-confounding covariates impact the causal inference, we first elucidate the performance of FAP methods in eliminating confounding bias based on the unified graphical framework depicted in Fig. \ref{UnifiedPers}. In the context of the observational data $\mathcal{D}$, we use the Average Treatment Effect (ATE) as a representative metric of the causal effect at the group level. The unadjusted $\hat{\text{ATE}}_{\mathcal{D}}$ is denoted in Eq.~\eqref{ATE_D}.
\begin{equation}
\label{ATE_D}
\hat{\text{ATE}}_{\mathcal{D}}=\frac{1}{N_{t}}\sum_{i:t_{i}=1}y_{i}-\frac{1}{N_{c}}\sum_{i:t_{i}=0}y_{i} = \bar{y_{t}} - \bar{y_{c}}
\end{equation}

Assuming a linear structural equation that relates the outcome variable to the treatment variable and confounders, expressed as $E[y_{i}(t)|C_{i} = c] = \gamma + \tau\cdot t + \beta\cdot c$, we can determine the true ATE value as $\tau$ ($\text{ATE}_{T} = \tau$). In addition, the estimation error for the unadjusted $\hat{\text{ATE}}_{\mathcal{D}}$ is described in Eq.~\eqref{Error_D} as $\Delta_{\mathcal{D}}$.
\begin{equation}
\label{Error_D}
\begin{aligned}
\Delta_{\mathcal{D}} &= \mathbb{E}[\hat{\text{ATE}}_{\mathcal{D}} - \text{ATE}_{T}] \\
&= \mathbb{E}\left[\bar{y_{t}}\right] - \mathbb{E}\left[\bar{y_{c}}\right] - \mathbb{E}[\text{ATE}_{T}]\\
&=  (\gamma + \tau + \beta\cdot \bar{C_{t}}) - (\gamma + \beta\cdot \bar{C_{c}}) - \tau\\
&= \beta\cdot (\bar{C_{t}} - \bar{C_{c}})
\end{aligned}
\end{equation}

Subsequently, the estimated ATE using the FAP method ($\hat{\text{ATE}}_{\text{FAP}}$), after adjusting for the proxy variable $P$, is expressed in Eq.~\eqref{ATE_FAP}.
\begin{equation}
\label{ATE_FAP}
\hat{\text{ATE}}_{\text{FAP}}=\sum_{j=1}^{J}q(j)\left[\bar{y_{t}}(j) - \bar{y_{c}}(j)\right]
\end{equation}
Where $j \in \{1,2,\cdots,J\}$ represent the $j$-th sub-block of the dataset $\mathcal{D}$. These $J$ sub-blocks are obtained by adjusting the proxy variable $P$. The $\bar{y_{t}}(j)$ and $\bar{y_{c}}(j)$ represent the mean of the outcome variable for the treatment and control group samples within the $j$-th sub-block, respectively. The proportion of samples in the $j$-th sub-block relative to the total population is denoted as $q(j)=N(j)/N$. Similarly, the estimation error $\Delta_{\text{FAP}}$ for $\hat{\text{ATE}}_{\text{FAP}}$ is illustrated as Eq.~\eqref{Error_FAP}. 
\begin{equation}
\label{Error_FAP}
\begin{aligned}
\Delta_{\text{FAP}} &= \mathbb{E}[\hat{\text{ATE}}_{\text{FAP}} - \text{ATE}_{T}] \\
&= \mathbb{E}\left[\sum_{j=1}^{J}q(j)\left[\bar{y_{t}}(j)\right]\right] - \mathbb{E}\left[\sum_{j=1}^{J}q(j)\left[\bar{y_{c}}(j)\right]\right] - \mathbb{E}[\text{ATE}_{T}]\\
&=  \left(\gamma + \tau + \beta\cdot \sum_{j=1}^{J}q(j)\bar{C_{t}}(j)\right) - \left(\gamma + \beta\cdot \sum_{j=1}^{J}q(j)\bar{C_{c}}(j)\right) - \tau\\
&= \beta\cdot \sum_{j=1}^{J}q(j)\left[\bar{C_{t}}(j) - \bar{C_{c}}(j)\right]
\end{aligned}
\end{equation}

In the ideal scenario, after applying the FAP method, there is no variance in confounders between the treatment and control groups within each sub-block of $\mathcal{D}$ (i.e., $\bar{C_{t}}(j) - \bar{C_{c}}(j) = \textbf{0}$), resulting in a subsequent reduction of $\Delta_{\text{FAP}}$ to zero. Therefore, it becomes evident that the FAP method reduces the initial estimation error $\Delta_{\mathcal{D}} = \beta\cdot (\bar{C_{t}} - \bar{C_{c}})$, which is directly estimated from $\mathcal{D}$, to $\Delta_{\text{FAP}}$ in Eq.~\eqref{Error_FAP}.
\subsection{The Perturbation for the Estimation Error $\Delta_{\text{FAP}}$ Instigated by Non-confounding Covariates}
In this subsection, we examine the variations in the estimated $\Delta_{\text{FAP}}$ when non-confounding variables, denoted as $U$, which include instrumental variables, adjustment variables, and so on, are included in the covariates. First, we derive the ATE estimation error $\Delta_{\text{FAP}}^{U}$  that includes $U$ in a similar manner as Eq.~\eqref{Error_FAP}, as shown in Eq.~\eqref{Error_FAP_U}.
%First, we incorporate non-confounding covariates into the structural equation for the outcome variable shown in Eq.~\eqref{struc_O1} and derive the ATE estimation error $\Delta_{\text{FAP}}^{U}$  that includes $U$ in a similar manner as Eq.~\eqref{Error_FAP}, as shown in Eq.~\eqref{Error_FAP_U}.
%\begin{equation}
%\label{struc_O1}
%E[y_{i}(t)|C_{i} = c, U_{i} = u] = \gamma + \tau\cdot t + \beta\cdot c  + \omega\cdot u
%\end{equation}
\begin{equation}
\label{Error_FAP_U}
\begin{aligned}
\Delta_{\text{FAP}}^{U} &= \mathbb{E}[\hat{\text {ATE}}_{\text{FAP}}^{U} - \text {ATE}_{T}] \\
&=  \left(\gamma + \tau + \beta\cdot \sum_{j=1}^{J}q(j)\bar{C}_{t}^{U}(j)\right) - \left(\gamma + \beta\cdot \sum_{j=1}^{J}q(j)\bar{C}_{c}^{U}(j)\right) - \tau\\
&= \beta\cdot \sum_{j=1}^{J}q(j)\left[\bar{C}_{t}^{U}(j) - \bar{C}_{c}^{U}(j)\right]
\end{aligned}
\end{equation}
In this context, $\bar{C}_{t}^{U}(j)$ and $\bar{C}_{c}^{U}(j)$ respectively represent the average confounding values within the $j$-th sub-block for the treatment and control group when $U$ is present. %Similarly, $\bar{U}_{t}(j)$ and $\bar{U}_{c}(j)$ respectively represent the values of non-confounding covariates within the $j$-th sub-block for the treatment and control group.

To clarify the relationship between $\Delta_{\text{FAP}}^{U}$ and $\Delta_{\text{FAP}}$, we assume that the proxy variable $P$ can be modeled by a linear structural equation. The influence of non-confounding variables is taken into account through two separate equations: Eq.~\eqref{struc_F1} represents the case when these variables are absent, while Eq.~\eqref{struc_F2} represents the scenario when they are present.
\begin{equation}
\label{struc_F1}
E[P_{i}|C_{i} = c] = \alpha_{0}\cdot c
\end{equation}
\begin{equation}
\label{struc_F2}
E[P_{i}^{U}|C_{i} = c, U_{i} = u] = \alpha_{1}\cdot c +\alpha_{2}\cdot u
\end{equation}
Based on Eq.~\eqref{struc_F2}, when adjusting for $P$, we assign the value $P(j)$ to $P$ within the $j$-th sub-block of $\mathcal{D}$. Consequently, the value of the confounding variable $C^{U}(j)$ within the $j$-th sub-block is determined as shown in Eq.~\eqref{conf_nc}, accounting for the presence of non-confounding variables. 
\begin{equation}
\label{conf_nc}
C^{U}(j|U = u) = \frac{P(j) -  \alpha_{2}\cdot u}{\alpha_{1}},\ \forall\ u
\end{equation}

\begin{figure}[ht]
\centering
\includegraphics[width=0.85\textwidth]{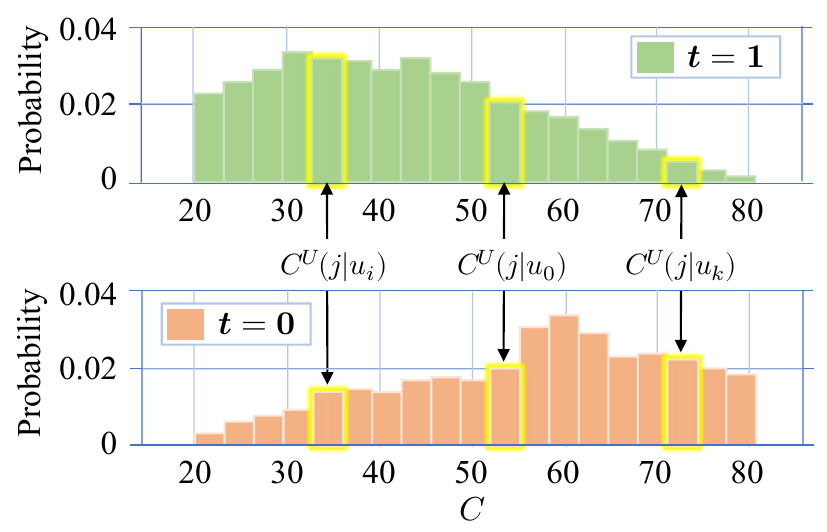}
\caption{A paradigmatic illustration of the perturbation caused by $U$ on the estimation error on ATE of the FAP. Consider a balanced distribution $P(j)$ with respect to the treatment variable $t$. The potential values of $C^{U}(j|U)$ include $C^{U}(j|u_{i})$, $C^{U}(j|u_{0})$, and $C^{U}(j|u_{k})$. Here, $C^{U}(j|u_{0})$ represents the equilibrium state, which corresponds to $C(j)$ in the absence of $U$.}
\label{U-Effect}
% \vspace{-0.3cm}
\end{figure}

Specifically, for any given $j$-th sub-block with a fixed value of $P(j)$, let $C(j)$ represents the confounding value when $U$ does not exist. In this case, there must exist an $nc_{0}$ such that $C(j) = C^{U}(j|u_{0})$. In other words, for the same $j$-th sub-block, when the non-confounding covariate takes the value $nc_{0}$, the value of $C^{U}(j|u_{0})$ is the same as the value of $C(j)$. On this basis, the specific implications of the non-confounding perturbation are illustrated in Fig. \ref{U-Effect}.

In Fig. \ref{U-Effect}, it can be observed that $C^{U}(j|u_{0})$ exhibits a state of approximate balance in relation to treatment $t$, similar to the situation in the absence of $U$. However, significant variations in the distribution of $C^{U}(j|u_{i})$ and $C^{U}(j|u_{k})$ exist among different treatment groups. Specifically, when $P(j)$ is balanced with respect to $t$, as indicated by Eq.~\eqref{struc_F1}, the distribution of $C(j)$ is also balanced with respect to $t$, expressed as $\mid\beta\cdot [\bar{C}_{t}(j) - \bar{C}_{c}(j)]\mid = 0$. However, if $U$ is mistakenly considered as the confounder, the balance of $C(j)$ is disrupted, denoted as $\mid\beta\cdot [\bar{C}_{t}^{U}(j) - \bar{C}_{c}^{U}(j)]\mid \geq 0$, leading to the inequalities expressed in Eq.~\eqref{inequalities}.
\begin{equation}
\label{inequalities}
\mid\bar{C}_{t}^{U}(j) - \bar{C}_{c}^{U}(j)\mid \geq \mid\bar{C}_{t}(j) - \bar{C}_{c}(j)\mid
\end{equation}

Equation~\eqref{inequalities} illustrates that for any sub-block $j$, even when the adjustment for the proxy variable $P$ is balanced with respect to the treatment $t$, there remains a difference in the distribution of confounders $C$ between the treatment and control groups when $U$ is present. Furthermore, this difference in distributions leads to a larger error $\Delta_{\text{FAP}}^{U}$ in estimating the ATE compared to $\Delta_{\text{FAP}}$.
\subsection{The Impacts of Specific Non-confounding Covariates on Causal Inference}
In the context of CIMs-POF, extending the proxy $P = f(C)$ to $P = f(C,U)$ denotes modeling the non-confounding covariates mistakenly as confounding variables. Consequently, the INCC problem translates into questioning the equivalence between adjusting for $f(C,U)$ and adjusting for $f(C)$. This equivalence is represented in Eq.~\eqref{SINCC}.
\begin{equation}
\label{SINCC}
\Delta_{\text{FAP}} \overset{\text{?}}{=} \Delta_{\text{FAP}}^{U} \Rightarrow
\sum_{j=1}^{J}q(j)\left[\bar{C_{t}}(j) - \bar{C_{c}}(j)\right] \overset{\text{?}}{=} \sum_{j=1}^{J}q(j)\left[\bar{C}_{t}^{U}(j) - \bar{C}_{c}^{U}(j)\right]
\end{equation}

We begin by presenting the conclusion regarding the issue of INCC: the equality holds in Eq.~\eqref{SINCC} when there is no causal relationship between the non-confounding covariates and the treatment variable. This means that the presence of these $U$ variables does not impact the adjustment for confounding bias on causal inference when this condition is met. To further support this conclusion, we partition the CIMs-POF based on the two distinct forms of the proxy variable $P$ and provide a proof for the aforementioned conclusion within each partition. 
\begin{itemize}
\item The first category of CIMs-POF utilizes a propensity score $P$ for $t$. These methods decompose the process of removing confounding bias into two sequential steps: (i) estimating the propensity score, and (ii) adjusting for confounding bias based on the estimated score. Representative methods in this category include Stratification, Matching, and IPW, etc. 
For these CIMs-POF, when there is no causal relationship between $U$ and $t$, the coefficient $\alpha_{2}$ of $U$ in Eq.~\eqref{struc_F2} is zero. As a result, $C^{U}(j) = P(j)/\alpha_{1}$ becomes independent of $U$, and the equality in Eq.~\eqref{SINCC} remains valid. 
\item In the second category of CIMs-POF, the proxy $P$ represents a balanced covariate representation in relation to $t$. These methods aim to eliminate confounding bias by optimizing the parameters of covariates with respect to $P$, such as $\alpha_{1}$ and $\alpha_{2}$ in Eq.~\eqref{struc_F2}. Representative methods belonging to this category include CFR, SITE, and others. 
When there is no causal relationship between $U$ and $t$, $U$ itself exhibits a balanced distribution relative to $t$. In such cases, although $\alpha_{2}$ may not be zero, the focus of these methods lies solely on optimizing $\alpha_{1}$ in Eq.~\eqref{struc_F2} to attain balance in $P$. Consequently, the optimization process is equivalent to confounders existing exclusively within the covariates.
\end{itemize}

While we have used ATE as a representation of causal effect in the preceding discussion, it is important to note that Individual Treatment Effect (ITE) estimation receives more attention in counterfactual inference tasks. In the aforementioned analysis, it is evident that there is no causal relationship between the adjustment variable $A$ and the treatment variable $t$. Therefore, the variable $A$ does not affect the confounding adjustment task. Moreover, since $A$ represents causal variable of the outcome variable $y$, the inclusion of $A$ in the covariates contributes to enhancing the precision of counterfactual inference and ITE estimation.

In summary, for the causal inference within POF, the best practice is to include only confounding variables during the confounding adjustment phase and to incorporate adjustment variables in the fitting of the counterfactual inference model. The inclusion of other non-confounding covariates can introduce additional inferential errors. This conclusion is further supported by the numerical experiments conducted.
\section{Experiments}
\subsection{Datasets}
In practice, researchers often encounter a situation where they can only observe the factual outcome of a particular treatment, while the corresponding counterfactual outcomes remain unknown. Additionally, identifying the precise components of the covariates in real-world datasets can be challenging. To address these limitations, previous studies have utilized synthetic or semi-synthetic datasets\cite{Hill2011Bayesian,dehejia1999causal,NIPS2017Latent,johansson2018learning}. In this paper, we aim to validate the INCC by leveraging a simulated dataset that incorporates labeled covariate components.

Based on the causal graph depicted in Fig. \ref{CASAUL}, the synthetic dataset $\mathcal{D}^{S}$ consists of three components: $\mathcal{D}^{S}= \{t,X,y\}$, where $X=\{I,C,A,M,Z,\textit{TI},\textit{YI}\,\}$. The variables $I$, $C$, and $A$ are exogenous variables and are randomly generated from a Normal distribution, as indicated in Eq.~\eqref{equ17}.
\begin{equation}
\label{equ17}
I, C, A \sim N(0,5)
\end{equation}

Subsequently, the treatment $t$ is generated randomly from a Bernoulli distribution, with the probabilities determined by $I$ and $C$, as illustrated in Eq.~\eqref{equ19}. Where, $f_{t}$ represents the sigmoid function, and $\epsilon_{t}$ follows a standard Normal distribution ($\epsilon_{t} \sim N(0,1)$).
\begin{gather}
t\sim B(P(t=1)) \nonumber\\
P(t=1) = 0.4\cdot f_{t}(C\cdot W_{C}^{t}) + 0.5\cdot f_{t}(I\cdot W_{I}^{t}) + 0.1\cdot f_{t}(\epsilon_{t}) \label{equ19}
\end{gather}

Next, the remaining non-root variables, namely $V = \{M, y, Z, \textit{TI}, \textit{YI}\,\}$ in Fig. \ref{CASAUL}, are sequentially generated based on their respective parent variables $Pa_{V}$, as depicted in Eq.~\eqref{non-root}. The specific parameter values are provided in Table~\ref{table:geda}. It is worth noting that all parameters $W$ follow a Uniform distribution within the range of $[2,5]$.
\begin{equation}
\label{non-root}
V = \sum_{Pa_{V}}f(Pa_{V}) + 0.1 * \epsilon_{V}, \epsilon_{V} \sim N(0,1)
\end{equation}
\begin{table}[h]
\centering
\caption{The Structural Equation of the Non-root Variables $V$ in $\mathcal{D}^{S}$.}
\label{table:geda}
%\resizebox{\linewidth}{!}{
\setlength{\tabcolsep}{0.14\textwidth}{ %set colWidth
\begin{tabular}{c|c}
\hline
$V$ & Equation $\sum_{Pa_{V}}f(Pa_{V})  + 0.1 * \epsilon_{V}$ ($f = sigmoid$) \\
 \hline
 $M$ & $f(t\cdot W_{t}^{M}) +  0.1 * \epsilon_{M}$ \\
 $y$ & $f(C\cdot W_{C}^{y}) + f(M\cdot W_{M}^{y}) + f(A\cdot W_{A}^{y}) +  0.1 * \epsilon_{y}$ \\
 $Z$ & $f(t\cdot W_{t}^{Z}) +  f(y\cdot W_{y}^{Z}) + 0.1 * \epsilon_{Z}$ \\
 $\textit{TI}$ & $f(t\cdot W_{t}^{\textit{TI}}) + 0.1 * \epsilon_{\textit{TI}}$ \\
 $\textit{YI}$ & $f(y\cdot W_{y}^{\textit{YI}}) +  0.1 * \epsilon_{\textit{YI}}$ \\
\hline
\end{tabular}
}
%}
\end{table}

The dataset $\mathcal{D}^{S}$ is comprised of a sample size of 20,000, generated according to the mechanism illustrated in Fig. \ref{CASAUL}. The covariates $X=\{I, C, A, M, Z, \textit{TI}, \textit{YI}\,\}$ have been explicitly defined, enabling a comprehensive analysis of the influence of different covariate components on the causal inference.
\subsection{Metrics} 
To illustrate the performance of CIMs-POF on the synthetic dataset $\mathcal{D}^{S}$, the errors of Individual Treatment Effect (ITE) and Average Treatment Effect (ATE) are used to assess inferential accuracy at the individual and group levels, respectively, as shown in Eq.~\eqref{ite}-\eqref{ate}. 
\begin{equation}
\label{ite}
\text{ITE}_{i} = y_{i}(t=1) - y_{i}(t=0) = t_{i}y_{i}^{f}-t_{i}y_{i}^{cf} + (1-t_{i})y_{i}^{cf} - (1-t_{i})y_{i}^{f}
\end{equation}
\begin{equation}
\label{ate}
\text{ATE} = \mathbb{E}\left[y(t=1)-y(t=0)\right] = \frac{1}{N}\sum_{i}^{N}\left[y_{i}(t=1)-y_{i}(t=0)\right]
\end{equation}

The evaluation metric precision in the estimation of heterogeneous effects (PEHE)\cite{Hill2011Bayesian} for the performance of ITE estimates, $\epsilon_{\text{PEHE}}$, as shown in Eq.~\eqref{equ21}, is computed by determining the mean square error between the predicted and actual ITE.
\begin{equation}
\label{equ21}
\epsilon_{\text{PEHE}} = \frac{1}{N}\sum_{i =1}^{N}\left(\left[y_{i}(t=1)-y_{i}(t=0)\right]-\left[\hat{y}_{i}(t=1)-\hat{y}_{i}(t=0)\right]\right)^{2}
\end{equation}
Where $y_{i}$ represents the ground-truth outcome, and $[y_{i}(t=1)-y_{i}(t=0)]$ reflects the actual value of the ITE. On the other hand, $\hat{y}_{i}$ denotes the predicted outcome, and $[\hat{y}_{i}(t=1)-\hat{y}_{i}(t=0)]$ represents the predicted value of the ITE.

In addition, $\epsilon_{\text{ATE}}$, absolute error of ATE is used to measure model performance at the group level, as shown in Eq.~\eqref{equ22}.
\begin{equation}
\label{equ22}
\epsilon_{\text{ATE}} = |\text{ATE} - \hat{\text{ATE}}|
\end{equation}
\subsection{Experimental Details}
In this section, we examine the consequences of misidentifying various non-confounding covariates as confounders for the causal inference. We evaluate the performance of various causal inference models, including Propensity Score-based Stratification (PSS)\cite{2015Principal}, Propensity Score-based Matching (PSM)\cite{2017Informative}, RCF\cite{2017Estimation}, IPW\cite{rosenbaum1983central}, CBPS and non-parametric CBPS (npCBPS)\cite{imai2014covariate}, Representation Space-based TARNet and CFR model\cite{shalit2017estimating}. These models are widely recognized and serve as representative examples of the CIMs-POF.

In order to demonstrate the impact of various non-confounding covariates, we consider the set of non-confounding covariates $U \in \{I,M,Z,A,\textit{TI},\textit{YI}\,\}$ as confounders in combination with the true confounders $C$, resulting in the covariate set $\{C,U\}$. For each combination of covariates $\{C,U\}$, all of the aforementioned models construct inferential models to estimate treatment effects and record the evaluation metrics $\epsilon_{\text{PEHE}}$ and $\epsilon_{\text{ATE}}$. The implementation details of each model are provided below.
\begin{itemize}
\item
For the propensity score-based models, the "glm" function in R language is used to estimate a logistic regression model for the combination of covariates $\{C,U\}$ with respect to the treatment $t$, represented as $\text{PS} = P(t=1|\{C,U\})$. Using the propensity scores, we calculate the group-level treatment effects, ATE, for the PSS, PSM, and IPW models based on Eq.~\eqref{ATE_FAP}, Eq.~\eqref{PSM}, and Eq.~\eqref{IPW}, respectively.
\begin{equation}
\label{PSM}
\hat{\text{ATE}}_{\text{PSM}}=\frac{1}{N}\sum_{i=1}^{N}\hat{y_{i}}(t=1) - \frac{1}{N}\sum_{i=1}^{N}\hat{y_{i}}(t=0)
\end{equation}
Where $\hat{y_{i}}(t)(t\in \{0,1\})$ is calculated by  Eq.~\eqref{PSMHATY}
\begin{equation}
\label{PSMHATY}
\hat{y_{i}}(t)=\begin{cases}\frac{1}{N(\mathcal{J}(i))} \sum_{l \in \mathcal{J}(i)} y_{l}, & t_{i}\neq t \\
y_{i}, & t_{i}=t\end{cases}
\end{equation}
In this context, $\mathcal{J}(i)$ represents the set of matched neighbors of the $i$-th sample in the opposite treatment group\cite{Austin2011An}, and $N(\mathcal{J}(i))$ denotes the number of samples within this neighbor set.
\begin{equation}
\label{IPW}
\hat{\text{ATE}}_{\text{IPW}} = \frac{1}{N}\sum_{i=1}^{N}t_{i}W_{i}y_{i} - \frac{1}{N}\sum_{i=1}^{N}(1-t_{i})W_{i}y_{i}
\end{equation}
Where $W_{i} = t_{i}/\text{PS}_{i}+(1-t_{i})/(1-\text{PS}_{i})$.
\item For the CBPS and npCBPS model, the sample weights were obtained by applying the "CBPS" package in R language and calculating the indicator ATE according to Eq.~\eqref{IPW}. The implementation of the RCF model is refered to the Python code provided in \cite{2017Estimation}. Similarly, for the TARNet and CFR model, we refer to the Python source code provided in \cite{shalit2017estimating}, which involved hyperparameters determined through a grid search method and remained consistent across all combinations of covariates. Since the TARNet and CFR model directly outputs counterfactual inference results, its estimated ITE and ATE indicators can be directly calculated from Eq.~\eqref{ite}-\eqref{ate}.
\item Except for the TARNet and CFR model, the PSS, PSM, RCF, IPW, CBPS and npCBPS models are implemented in a two-step process to address the tasks of confounding elimination and counterfactual inference. In particular, the "gbm" package in R language is consistently utilized to train a nonlinear gradient boosting model (GBM) for conducting counterfactual inference. The optimal number of trees is determined using the built-in optimization function "gbm.perf".
\item For the evaluation of ITE metrics, the PSS model begins by constructing a GBM within each stratum after stratification based on the propensity score. Subsequently, a weighted average of the ITE is computed for all stratums; The PSM model first constructs a matched dataset based on the propensity score and then builds a GBM on the matched dataset to calculate the ITE metrics; On the other hand, the RCF, IPW, CBPS and npCBPS approaches develop weighted GBM using the respective learned sample weights and subsequently calculate the ITE metrics.
\end{itemize}

\subsection{Results}
%Table \ref{ATEPEHE} presents the individual and group-level inference errors, denoted as $\epsilon_{PEHE}$ and $\epsilon_{ATE}$ respectively, for the CIMs-POF on the synthetic dataset $\mathcal{D}^{S}$ across different combinations of covariates $\{C,U\}$. The combinations of covariates $\{C,U\}$ that exhibit the best and worst inference performance for each CIM-B-POF are highlighted in blue and red, respectively. Based on these results, we make the following observations and analyses:
Table \ref{ATE-M} presents the group level inference errors, denoted as $\epsilon_{\text{ATE}}$, for the CIMs-POF on the synthetic dataset across different combinations of covariates $\{C,U\}$. The combinations of covariates $\{C,U\}$ that exhibit the best and worst inference performance for each CIM-B-POF are highlighted in blue and red, respectively. Based on these results, we make the following observations and analyses:
\begin{table}[!h]
\centering
\caption{Performance of $\epsilon_{\text{ATE}}$ estimation of the CIMs-POF on synthetic datasets under various $\{C,U\}$. Where bold blue marks covariate combinations for optimal performance under each model and bold red vice versa.}
\label{ATE-M}
%\resizebox{\linewidth}{!}{
\setlength{\tabcolsep}{0.022\textwidth}{ %set colWidth
\begin{tabular}{l@{}c|c|c|c|c|c|c|c}
\hline
\multirow{2}{*}{\bm{$\{C,U\}$}} &  \multicolumn{8}{c}{{\bfseries Metrics: \bm{$\epsilon_{\text{ATE}}$}}}\\
\cline{2-9}
& {\bfseries PSS} & {\bfseries PSM} & {\bfseries RCF} &  {\bfseries IPW}  & {\bfseries CBPS} & {\bfseries npCBPS} & {\bfseries TARNet} & {\bfseries CFR} \\
\hline
\multicolumn{1}{l|}{$\{C\}$} & \textbf{\textcolor{blue}{0.06}} & \textbf{\textcolor{blue}{0.03}} & \textbf{\textcolor{blue}{0.05}} & \textbf{\textcolor{blue}{0.11}} & \textbf{\textcolor{blue}{0.19}} & \textbf{\textcolor{blue}{0.08}} & 0.06 & 0.17\\ 
\multicolumn{1}{l|}{$\{C,I\}$}  & 0.19 & 0.19 & 0.07 & 0.17 & 0.21 & 0.11 & 0.07 & 0.26\\
\multicolumn{1}{l|}{$\{C,A\}$}  & \textbf{\textcolor{blue}{0.06}} & \textbf{\textcolor{blue}{0.03}} & \textbf{\textcolor{blue}{0.05}} & \textbf{\textcolor{blue}{0.11}} & \textbf{\textcolor{blue}{0.19}} & \textbf{\textcolor{blue}{0.08}} & \textbf{\textcolor{blue}{0.04}} & \textbf{\textcolor{blue}{0.09}}\\
\multicolumn{1}{l|}{$\{C,\textit{YI}\,\}$}  & \textbf{\textcolor{blue}{0.06}} & \textbf{\textcolor{blue}{0.03}} & 0.06 & \textbf{\textcolor{blue}{0.11}} & \textbf{\textcolor{blue}{0.19}} & \textbf{\textcolor{blue}{0.08}} & \textbf{\textcolor{blue}{0.04}} & 0.10\\
\multicolumn{1}{l|}{$\{C,M\}$}  & 0.40 & 0.58 & 0.58 & 0.51 & 0.54 & 0.12 & 0.14 & 0.40\\
\multicolumn{1}{l|}{$\{C,\textit{TI}\,\}$}  & 0.16 & 0.32 & 0.43 & 0.36 & 0.33 & 0.12 & 0.12 & 0.43\\
\multicolumn{1}{l|}{$\{C,Z\}$}  & \textbf{\textcolor{red}{1.20}} & \textbf{\textcolor{red}{0.39}} & \textbf{\textcolor{red}{0.96}} & \textbf{\textcolor{red}{2.67}} & \textbf{\textcolor{red}{2.21}} & \textbf{\textcolor{red}{0.43}} & \textbf{\textcolor{red}{1.27}} & \textbf{\textcolor{red}{0.68}}\\
\hline
\end{tabular}
}
%}
\end{table}
\begin{itemize}
%\textcolor{green}{ }
\item In terms of the metric $\epsilon_{\text{ATE}}$, the covariate combination $\{C\}$ exhibits identical performance to $\{C,A\}$, except for the TARNet and CFR model. This observation highlights that the adjustment variable $A$ does not impact confounding elimination. The divergent performance of the TARNet and CFR method can be attributed to the fact that its metrics, $\epsilon_{\text{ATE}}$ and $\epsilon_{\text{PEHE}}$, are generated by a single model, whereas the $\epsilon_{\text{ATE}}$ values of other methods are calculated directly after confounding elimination, without modeling the counterfactual inference model.
\item The combination of covariates $\{C,\textit{YI}\,\}$ performs similarly to $\{C,A\}$, primarily due to the presence of a distant causal path between the treatment variable $t$ and $\textit{YI}$, namely $t \rightarrow M \rightarrow y \rightarrow \textit{YI}$. This is supported by the small correlation observed between $t$ and $\textit{YI}$ in the synthetic dataset $\mathcal{D}^{S}$. Furthermore, there exists a reverse causal relationship between $\textit{YI}$ and the outcome variable $y$, resulting in its approximate inferential accuracy with the variable $A$. However, it is important to note that this observation may not always be valid, as its validity depends on the extent of correlation between the $t$, the $y$, and the $\textit{YI}$ in the actual dataset.
\end{itemize}

In addition, we record the inferential performance ($\epsilon_{\text{PEHE}}$) of CIMs-POF for different combinations $\{C,U\}$ at the individual level in Table \ref{PEHE-M}. Based on Table \ref{PEHE-M} the following analyses can be drawn:
\begin{itemize}
%\textcolor{green}{ }
\item For both individual and group-level inferred errors $\epsilon_{\text{PEHE}}$ and $\epsilon_{\text{ATE}}$, the combination of covariates $\{C,A\}$ yielded the optimal results across all models. This finding is consistent with the theoretical analysis, suggesting that during the confounding adjustment phase, the covariates should ideally include only the confounding variable. Additionally, the adjustment variable $A$ does not impact this step, but it does contribute to a more precise prediction of the counterfactual outcomes.
\item Among all the combinations of covariates, $\{C,Z\}$ exhibited the worst inferential performance at both the individual and group levels. The theoretical foundation of this phenomenon is the introduction of additional correlations between $t$ and $y$ within $j$-th sub-block of $\mathcal{D}$ when adjusting for the colliders $Z$, as illustrated in Eq.~\eqref{colliders}.
\begin{equation}
\label{colliders}
E[y_{i}(t)|C_{i} = c] = \gamma + \left(\tau + \tau^{'}\left(j\right)\right)\cdot t + \beta\cdot c
\end{equation}
Based on Eq.~\eqref{colliders}, the estimation error transitions from 
$\Delta_{\text{FAP}}^{U}$ shown in Eq.~\eqref{Error_FAP_U} to $\Delta_{\text{FAP}}^{Z}$ shown in Eq.~\eqref{FAP_Z}.
\begin{equation}
\label{FAP_Z}
\Delta_{\text{FAP}}^{Z} = \beta\cdot \sum_{j=1}^{J}q(j)\left[\bar{C}_{t}^{Z}(j) - \bar{C}_{c}^{Z}(j) + \tau^{'}(j)\right]
\end{equation}
Where $Z$ retains its arbitrariness and $\beta\cdot \sum_{j=1}^{J}q(j)\tau^{'}(j)$ represents the additional bias introduced relative to other $U$. Consequently, the CIMs-POF exhibit the worst inferential performance when the colliders are mistakenly treated as the confounders.
\end{itemize}
\begin{table}[!ht]
\centering
\caption{Performance of $\epsilon_{\text{PEHE}}$ estimation of the CIMs-POF on synthetic datasets under various $\{C,U\}$. Where bold blue marks covariate combinations for optimal performance under each model and bold red vice versa.}
\label{PEHE-M}
%\resizebox{\linewidth}{!}{
\setlength{\tabcolsep}{0.022\textwidth}{ %set colWidth
\begin{tabular}{l@{}c|c|c|c|c|c|c|c}
\hline
\multirow{3}{*}{\bm{$\{C,U\}$}} &  \multicolumn{8}{c}{{\bfseries Metrics: \bm{$\epsilon_{\text{PEHE}}$}}}\\
\cline{2-9}
& {\bfseries PSS} & {\bfseries PSM} & {\bfseries RCF} &  {\bfseries IPW}  & {\bfseries CBPS} & {\bfseries npCBPS} & {\bfseries TARNet} & {\bfseries CFR} \\
\hline
\multicolumn{1}{l|}{$\{C\}$} & 1.25 & 0.69 & 1.26 & 1.25 & 1.25 & 1.26 & 0.11 & 0.22 \\ 
\multicolumn{1}{l|}{$\{C,I\}$}  & 1.29 & 0.77 & 1.28 & 1.25 & 1.26 & 1.28 & 0.13 & 0.34 \\
\multicolumn{1}{l|}{$\{C,A\}$}  & \textbf{\textcolor{blue}{1.04}} & \textbf{\textcolor{blue}{0.67}} & \textbf{\textcolor{blue}{1.02}} & \textbf{\textcolor{blue}{1.02}} & \textbf{\textcolor{blue}{1.02}} & \textbf{\textcolor{blue}{1.02}} & \textbf{\textcolor{blue}{0.10}} & \textbf{\textcolor{blue}{0.16}} \\
\multicolumn{1}{l|}{$\{C,\textit{YI}\,\}$}  & 1.05 & 0.80 & \textbf{\textcolor{blue}{1.02}} & \textbf{\textcolor{blue}{1.02}} & \textbf{\textcolor{blue}{1.02}} & 1.03 & 0.11 & \textbf{\textcolor{blue}{0.16}} \\
\multicolumn{1}{l|}{$\{C,M\}$}  & 1.52 & 1.16 & 1.42 & 1.30 & 1.34 & 1.35 & 0.18 & 0.42 \\
\multicolumn{1}{l|}{$\{C,\textit{TI}\,\}$}  & 1.46 & 0.91 & 1.41 & 1.32 & 1.37 & 1.48 & 0.16 & 0.45 \\
\multicolumn{1}{l|}{$\{C,Z\}$}  & \textbf{\textcolor{red}{3.06}} & \textbf{\textcolor{red}{1.73}} & \textbf{\textcolor{red}{2.92}} & \textbf{\textcolor{red}{3.79}} & \textbf{\textcolor{red}{3.10}} & \textbf{\textcolor{red}{3.03}} & \textbf{\textcolor{red}{1.31}} & \textbf{\textcolor{red}{0.83}} \\
\hline
\end{tabular}
}
%}
\end{table}
%\begin{table}[!t]
%\centering
%\caption{PEHE.}
%\label{table:PEHE}
%\resizebox{\linewidth}{!}{
%\begin{tabular}{l|c|c|c|c|c}
%\hline
 %$\{X\}$ & Stratification &  Matching & IPW & CBPS & CFR\\
% \hline
% $\{C\}$ & 1.25 & 0.69 & 1.25 & 1.25 & 0.22 \\ 
% $\{C,I\}$ & 1.29 & 0.77 & 1.25 & 1.26 & 0.34 \\
% $\{C,A\}$ & 1.04 & 0.67 & 1.02 & 1.02 & 0.16 \\
% $\{C,YI\}$ & 1.05 & 0.80 & 1.02 & 1.02 & 0.16 \\
% $\{C,M\}$ & 1.52 & 1.16 & 1.30 & 1.34 & 0.42 \\
% $\{C,TI\}$ & 1.46 & 0.91 & 1.32 & 1.37 & 0.45 \\
% $\{C,Z\}$ & 3.06 & 0.73 & 3.79 & 3.10 & 0.83 \\
%\hline
%\end{tabular}
%}
%\end{table}

%\begin{table}[!t]
%\centering
%\caption{ate.}
%\label{table:ate}
%\resizebox{\linewidth}{!}{
%\begin{tabular}{c|c|c|c|c|c}
%\hline
% $\{X\}$ &  Stratification &  Matching & IPW & CBPS & CFR\\
% \hline
% $\{C\}$ & 0.06 & 0.03 & 0.11 & 0.19 & 0.17 \\ 
% $\{C,I\}$ & 0.19 & 0.19 & 0.17 & 0.21 & 0.26 \\
% $\{C,A\}$ & 0.06 & 0.03 & 0.11 & 0.19 & 0.09 \\
% $\{C,YI\}$ & 0.06 & 0.03 & 0.11 & 0.19 & 0.10 \\
% $\{C,M\}$ & 0.40 & 0.58 & 0.51 & 0.54 & 0.40 \\
% $\{C,TI\}$ & 0.16 & 0.32 & 0.36 & 0.33 & 0.43 \\
% $\{C,Z\}$ & 1.20 & 0.39 & 2.67 & 2.21 & 0.68 \\
%\hline
%\end{tabular}
%}
%\end{table}
\section{Conclusion}
The CIMs-POF are highly regarded in the field of causal inference, which plays a crucial role in domains such as economics, healthcare, and epidemiology.
Most CIMs-POF default to assuming that all covariates are confounders, which is referred to as the assumption of Confounding Covariates. However, this assumption is challenging to fulfill in practice, particularly when dealing with high-dimensional covariates. The influence mechanism for treating non-confounding covariates as confounders remains unclear, raising concerns about the applicability of the causal inference.
In this paper, we propose a unified graphical framework for CIMs-POF and present the instances of the framework for various representative CIMs-POF. This framework facilitates a cohesive understanding of the shared operational principles among different CIMs-POF.
Based on graphical framework, we quantitatively assess the impact on causal inference when non-confounding variables, such as instrumental variables, mediators, colliders, adjustment variables, treatment-influenced variables, and outcome-influenced variables, are mistakenly considered as confounding variables. This quantification is achieved through a rigorous theoretical analysis.
Furthermore, the experimental findings on the synthetic dataset align consistently with the theoretical analysis, indicating that optimal performance of the CIMs-POF in eliminating confounding bias and conducting counterfactual inference can be achieved by including only confounders and adjustment variables in the covariates.
In the future, there is a pressing need to focus on designing algorithms that can systematically identify specific components of covariates. 
%\appendix
%\section{Derivation of the optimization objectives for the VLUCI}
%\subsection{Derivation of the optimization objectives for the LCDL}


\bibliographystyle{cas-model2-names}
% \bibliographystyle{model1-num}

\bibliography{mybib}

% \newpage
% \vskip3pt

\end{document}